%Paper: hep-ph/9209299
%From: cliff@hep.Physics.McGill.CA (Cliff Burgess)
%Date: Wed, 30 Sep 92 17:51:30 -0400

\input harvmac.tex

\def\ell{p}
\def\cc{\hbox{c.c.}}
\def\sss{\scriptscriptstyle}
\def\G{\Gamma}
\def\pf{p_{\sss F}}
\def\GF{G_{\sss F}}
\def\psir{\Psi_{\sss R}}
\def\psil{\Psi_{\sss L}}
\def\bb{\beta\beta}
\def\bbtn{\bb_{2\nu}}
\def\bbm{\bb_{M}}
\def\bbom{\bb_{o\sss M}}
\def\bbcm{\bb_{c\sss M}}
\def\bbzn{\bb_{0\nu}}
\def\gnn{g_{\nu_e}}

\def\etal{{\it et al.}}

\def\Pr{P_{\sss R}}
\def\Pl{P_{\sss L}}
\def\M{{\cal M}}
\def\hti{\tilde{h}}
\def\VEV#1{\left\langle #1\right\rangle}
\def\VEVs#1{\left|\left\langle #1\right\rangle\right|^2}
\def\gtwid{\mathrel{\raise.3ex\hbox{$>$\kern-.75em\lower1ex\hbox{$\sim$}}}}
\def\ltwid{\mathrel{\raise.3ex\hbox{$<$\kern-.75em\lower1ex\hbox{$\sim$}}}}

\def\frac#1#2{{\scriptstyle{#1 \over #2}}}
\def\slp{{\raise.15ex\hbox{$/$}\kern-.57em\hbox{$\partial$}}}
\def\gv{g_{\sss V}}
\def\ga{g_{\sss A}}

\def\ellsl{{\raise.15ex\hbox{$/$}\kern-.57em\hbox{$\ell$}}}
\def\psl{{\raise.15ex\hbox{$/$}\kern-.57em\hbox{$p$}}}
\def\qsl{{\raise.15ex\hbox{$/$}\kern-.57em\hbox{$q$}}}

\def\Nd{$^{150}$Nd}
\def\Mo{$^{100}$Mo}
\def\Se{$^{82}$Se}
\def\Nd{$^{150}$Nd}
\def\U{$^{238}$U}
\def\Ge{$^{76}$Ge}
\def\Te#1{$^{#1}$Te}
\def\hbr{\hat{\bf r}}
\def\sm{{\bf\sigma}_m}
\def\sn{{\bf\sigma}_n}
\def\Dm{{\bf D}_m}
\def\Dn{{\bf D}_n}

\def\SN#1#2{#1\times 10^{#2}}
\def\SNt#1#2{$#1\times 10^{#2}$}
\def\pr#1{Phys.~Rev.~{\bf #1}}
\def\np#1{Nucl.~Phys.~{\bf #1}}
\def\pl#1{Phys.~Lett.~{\bf #1}}
\def\prl#1{Phys.~Rev.~Lett.~{\bf #1}}
\def\ie{{\it i.e.}}

\def\ol#1{\overline{#1}}
\def\geff{g_{\rm eff}}

\def\Sca{{\cal A}}
\def\Scl{{\cal L}}
\def\Scm{{\cal M}}
\def\Scu{{\cal U}}
\def\Scv{{\cal V}}
\def\hf{{1\over 2}}
\def\keV{{\rm \ keV}}
\def\MeV{{\rm \ MeV}}

\def\roughlyup#1{\mathrel{\raise.3ex\hbox{$\sim$\kern-.75em\lower1ex\hbox{$#1$}
}}}
\def\roughlydown#1{\mathrel{\raise.3ex\hbox{$#1$\kern-.75em\lower1ex\hbox{$\sim$
}}}}
\def\bra{\langle}
\def\ket{\rangle}
\def\lsim{\roughlydown<}
\def\gsim{\roughlydown>}
\def\simeq{\roughlyup-}
\def\mnu#1{m_{\nu_#1}}

\tolerance=10000
\hfuzz=5pt
\overfullrule=0pt

\Title{\vbox{\baselineskip12pt\hbox{McGill/92-22}\hbox{hep-ph/9207207}
\hbox{rev. August 1992}}}
{\vbox{\centerline{Majorons without Majorana Masses}
\centerline{}
\centerline{and Neutrinoless Double Beta Decay}}}
\centerline{C.P.~Burgess and J.M.~Cline$^*$}\footnote{}{$^*$ Present address:
Theoretical Physics Institute, University of Minnesota, Minneapolis, MN 55455,
USA}
\centerline{McGill University}
\centerline{3600 University Street}
\centerline{Montr\'eal, Qu\'ebec, Canada H3A 2T8}
\vskip .3in
We explain excess events near the endpoints of the double beta
decay ($\bb$) spectra of several elements, using the neutrinoless emission of
massless Goldstone bosons.  Models with scalars carrying lepton number $-2$ are
proposed for this purpose so that ordinary neutrinoless $\bb$ is forbidden, and
we can raise the scale of global symmetry breaking above the 10 keV scale
needed for observable emission of conventional Majorons in $\bb$.  The electron
spectrum has a different shape, and the rate depends on different nuclear
matrix elements, than for the emission of ordinary Majorons.
\Date{}
One of the deep questions of particle physics is whether there exist any
fundamental particles of spin zero. A candidate is the Majoron, the massless
Goldstone boson that would exist if lepton number ($L$) were spontaneously
broken \ref\CMP{Y.~Chikashige, R.N.~Mohapatra and R.D.~Peccei, \pl{99B} (1981)
411.}. Majorons would be hard to detect since they couple directly only to
neutrinos, with a strength proportional to the neutrino masses divided by the
$L$-breaking scale $v$:
     \eqn\coupling{g_\nu = m_\nu/v.}

If $g_\nu$ were large enough, Majorons could be seen in double beta decay
\ref\Doi{M.~Doi, T.~Kotani and E.~Takasugi, \pr{D37} (1988)
2572.}\ref\georgi{H.M.~Georgi, S.L.~Glashow and S.~Nussinov, \np{B193} (1981)
297.}, a rare process now observed in seven elements.  Besides the usual
neutrino-emitting mode $\bbtn$, a Majoron could be emitted through annihilation
of the virtual neutrinos, $\bbm$.  The $\bbm$ and $\bbtn$ signals are
distinguishable since for $\bbm$ the decay energy is shared among fewer
particles, skewing its electron spectrum toward higher energies.

In fact, a mysterious excess of high-energy electrons is seen in the $\bb$
spectra for several elements.  Such an observation was first made in 1987 for
$^{76}$Ge $\to^{76}$Se $+2e^-$ by Avignone \etal\ \ref\Avignone{F.T.~Avignone
III \etal, in {\it Neutrino Masses and Neutrino Astrophysics,} proceedings of
the IV Telemark Conference, Ashland, Wisconsin, 1987, edited by V.~Barger,
F.~Halzen, M.~Marshak and K.~Olive  (World Scientific, Sinagpore, 1987),
p.~248.}, although they, and other groups, subsequently excluded a signal at
the
original level \ref\nomajoron{P.~Fisher \etal, \pl{B192} (1987) 460;
D.O.~Caldwell \etal, \prl{59} (1987) 1649.}.  Now the UC Irvine group also
finds
excess numbers of electrons near but below the endpoints for \Mo, \Se\ and \Nd,
with a statistical significance of 5$\sigma$ \ref\Moe{M.~Moe, M. Nelson, M.
Vient and S. Elliott, preprint UCI-NEUTRINO 92-1.}. Such events also persist in
${}^{76}$Ge \ref\priv{F.T. Avignone, private communication}, at approximately a
tenth of the original rate.

Formerly all these $\bb$ measurements were compared with the Gelmini-Roncadelli
(GR) model \ref\GR{G.B.~Gelmini and M.~Roncadelli, \pl{99B}  (1981) 411.}
\georgi, which however has been ruled out by LEP's bounds on the invisible Z
width.  In this model lepton number was spontaneously  broken by an
electroweak-triplet Higgs field, resulting in both a Majorana mass and a direct
coupling to the Majoron for $\nu_e$, as in eq.~\coupling.   The presently
observed excess events could then be explained by $\bbm$ if  $\gnn\sim
\SN{1}{-4}$, and the absence of $\bbzn$ decay --- which would appear as a line
at the endpoint in the sum-energy spectrum --- requires $\mnu{e} \lsim 1$ eV.
Eq.~\coupling\ then implies that the triplet VEV must be unnaturally small,
   \eqn\bound{ v < 10 \keV.}
(A similar bound follows from astrophysical considerations.)

With the demise of the GR model, it appeared that there existed no models
capable of predicting $\bbm$ at an observable rate. In this Letter we propose
an alternative broad class of Majoron models which might explain the excess
events, while still preserving the agreement between theory and geochemical
experiments  for the $\bb$ decay rate of \Te{128}, \Te{130}\ and \U\
\ref\uranium{We have been informed that the discrepancy between the \U\
observations of Turkevich, Economou and Cowan, (\prl{67} (1991) 3211) and the
calculations of Staudt, Muto and Klapdor-Kleingrothaus (Europhys.~Lett., {\bf
13} (1990) 31) have now been resolved.}.

We start by describing some generic features of $\bbm$ decays, and then turn to
model-specific issues. First, agreement with the $Z$-width measurement rules
out a direct Majoron coupling to the $Z$ boson. Lepton number breaking,
if it occurs, must come from an electroweak-singlet field.  Majorons therefore
cannot have renormalizable couplings to $\nu_e$; instead, $\bbm$ proceeds
through mixing of $\nu_e$ with sterile neutrinos that couple to Majorons.

Next one must explain how $\bbm$ but not $\bbzn$ events are seen, since
lepton-breaking generically gives both effects.  The GR model did so by
requiring a small lepton-number breaking scale as in eq.~\bound, and the same
expedient is used by alternate models  where electroweak-singlets rather than
triplets do the lepton-breaking. But once such a small scale is introduced by
hand it is unnecessary for the scalar emitted in $\bbm$ to be a Goldstone
boson; it could just as well be given a {\it mass} of 10 keV.  We refer to
these
as `ordinary Majorons' (and denote their emission by $\bbom$) since they share
many features (electron spectra, etc.) of the GR model. These models all suffer
from a hierarchy problem since the lepton-breaking VEV is not stable under
renormalization and so must be artificially fine-tuned \ref\valle{It has
recently been claimed by Z.G.~Berezhiani, A.Yu.~Smirnov and J.W.F.~Valle,
(preprint FTUV/92-20,  IFIC/92-21, LMU-05/92) that the required Majoron
coupling, $g_{\sss M}\sim 10^{-4}$, is sufficiently small to allow such a
hierarchy below the weak scale. This argument overlooks the fact that in these
models the effective Majoron coupling measured in $\bbm$ decay is given by $g_M
\sim g \theta^2$, where the small mixing angle, $\theta$, is bounded by
oscillation and neutrino decay experiments to be very small. Since the coupling
that controls the hierarchy problem is $g$, and {\it not} $g_M$, it is
typically
not small: $g\sim O(1)$.}.

Here we introduce a class of models that do not require such a small scale,
inspired by realizing that the smallness of $v$ in eq.~\bound\ comes from the
bound on $\bbzn$ decay.  This restriction disappears if a lepton number, $L$,
carried by electrons, remains unbroken, so that $\bbzn$ is completely
forbidden.  Then $\bbm$ occurs only if the scalar particle itself is `charged,'
\ie, it carries $L(\varphi)=-2$, in units where  $L(e^-)=+1$.  Fine-tuning of
the scalar mass is avoided if the scalar is a Goldstone boson, a `charged
Majoron.'  All of this can be realized in a simple way, using for example a
global $SU(2)_s\times U(1)_{L'}$ symmetry broken down to $U(1)_L$. (The
$U(1)_{L'}$ factor exists so that  Majoron couplings to ordinary charged
leptons can be avoided.)  Two of the resulting Goldstone bosons then carry the
unbroken charge, just as the would-be Goldstone bosons that make up the
longitudinal $W^\pm$'s carry electric charge in the Standard Model. We call all
such models `charged' Majoron models and denote the resulting process $\bbcm$.

In what follows we compare charged and ordinary Majoron models with each
another and the data.  We will: ($i$) show how the models can be experimentally
distinguished by the predicted shape of the sum electron spectrum, and ($ii$)
find the necessary conditions for getting a large enough  rate of $\bbm$.

{\it i.~The electron spectrum.}  The decay rates for $\bbtn$, $\bbom$ and
$\bbcm$ all have the form
 \eqn\genericrate{ d\G(\bb) = {\GF^4 \over 4\pi^3} \; \left|
   \Sca(\bb) \right|^2 d\Omega_n,}
where $\GF$ is the Fermi constant, $\Sca$ is a matrix element (more about
which later), and $d\Omega_n$ is the phase space, a function of
the momenta and energies of the two outgoing electrons,
   \eqn\phsp{d\Omega_n = {1\over 64\pi^2}(Q - \epsilon_1-\epsilon_2)^n
   \prod_{k=1}^2 p_k \epsilon_k F(\epsilon_k) d\epsilon_k.}
$F(\epsilon)$ is the Fermi function, equal to 1 when the nuclear charge
vanishes, and $Q\simeq 1-2 \, \MeV$ is the endpoint energy for the final-state
electrons.

The integer $n$ in eq.~\phsp\ depends on the decay channel: $n=5$ for $\bbtn$,
$3$ for $\bbcm$, and $1$ for $\bbom$. $n$ controls the shape of the electron
energy spectrum, because for these decays all lepton energies are $\ll \pf\sim
100$ MeV, the nuclear scale that determines the matrix element $\Sca$. Thus it
is a good approximation --- relative error of order $Q/\pf\sim 1\%$ --- to
neglect $\Sca$'s dependence on the outgoing lepton energies. In particular,
near the endpoint the electron spectrum in $\bbcm$ goes like $(Q-E)^3$ as $E$,
the sum of electron energies, approaches $Q$.  This is intermediate between
$(Q-E)$ for $\bbom$, and $(Q-E)^5$ for $\bbtn$. The spectra are plotted in
Fig.~1. The distinction between $\bbcm$ and $\bbom$ is one of our main results.

\midinsert
\vskip 2in
Figure 1. The electron sum-energy spectrum for $\bbtn$ (dashed), $\bbcm$
(solid)
and $\bbom$ (dot-dashed) decays.
\bigskip
\endinsert

We emphasize that the spectral shape difference characterizes these two classes
of models, independently of any model-specific details.  This traces to the
vanishing of $\bbcm$ amplitudes at zero Majoron momentum, $q = 0$, due to the
fact that Goldstone bosons are always derivatively coupled. The puzzle is why
$\bbom$ amplitudes do not also vanish at $q=0$.  There one finds that graphs
with Majoron Brehmsstrahlung off the electron lines survive as $q\to 0$ because
the internal electron propagator diverges in this limit,  leaving a finite
result. The same cannot happen for $\bbcm$ because the electron-Majoron
coupling
is forbidden by the unbroken lepton symmetry in these models.

{\it ii.~The decay rate.} Computing the absolute rate requires knowing the
matrix element, $\Sca$, of eq.~\genericrate. Consider arbitrary Yukawa
couplings between a set of neutrinos and a complex scalar field, $\varphi$,
   \eqn\yukawa{\Scl_M = - \hf\;     \ol{\nu}_i (a_{ij} \Pl  + b_{ij}  \Pr)
   \; \nu_j \; \varphi^* + c.c. }
Goldstone boson couplings are included here as a special case: if
$f$ is the decay constant, $X_{ij}$ the conserved charge and $m_{ij}$ the
left-handed neutrino mass matrix, then $a_{ij}$ takes the form
$a = -i (X^T m + m X)/f$, and similarly for $b_{ij}$. Real scalar fields are
also easily incorporated  by taking $b = a^*$ in all results. For example the
GR Majoron coupling arising  from a triplet interaction $\Scl = \hf g \;
\ol{L}_e \Pl \tau_a L_e \; \Delta^a + \cc$ would give $b^*_{ee} = a_{ee} = i
g$.

To lowest (zeroeth) order in the scalar energy we get eq.~\genericrate\ with
$n=1$ and
  \eqn\ordinaryrate{ \Sca(\bbom) = \sum_{ij} V_{ei} V_{ej}
  \int {d^4\ell \over (2\pi)^4} \; {(4 w_1 - w_2 + \ell^2 w_5)
  \; (a_{ij} \mnu{i} \mnu{j} - \ell^2 b_{ij}) \over
  (\ell^2 + \mnu{i}^2 -i\varepsilon )(\ell^2 + \mnu{j}^2 -i\varepsilon)}. }
$V_{ei}$ is the mixing matrix element for finding the mass eigenstate
$\nu_i$ in an electron charged-current weak interaction. The form factors
$w_a$ parametrize the matrix element of the hadronic weak currents
between initial and final nuclei $N$ and $N'$,
  \eqn\matrixelement{ \eqalign{ W_{\alpha\beta} &\equiv \int d^4x\;\bra N'|T^*
     \left[ J_\alpha(x) J_\beta(0) \right] | N\ket \; e^{i\ell x} \cr
     &= w_1 \; \eta_{\alpha\beta} + w_2 \; u_\alpha u_\beta + w_3 \;
     (\ell_\alpha u_\beta - \ell_\beta u_\alpha) + \cr
     & \qquad + w_4 \; \epsilon_{\alpha\beta\sigma\rho}
     u^\sigma \ell^\rho + w_5 \; \ell_\alpha \ell_\beta, \cr}}
and are functions of the invariants $\ell^2$ and $u\!\cdot\!\ell$. $u_\alpha$
is the four-velocity of the initial and final nuclei, which are equal since we
neglect all dependence in $W_{\alpha\beta}$ on the final-state momenta.

Although eq.~\ordinaryrate\ suffices for $\bbom$ it happens that for $\bbcm$
this lowest-order result vanishes, making it necessary to work to next-higher
order in the outgoing Majoron momentum, and leading to the difference in
electron spectra (above).  The result  has the form of eq.~\genericrate, but
with $n=3$ and   $|\Sca(\bbcm)|^2 = |\Scu(\bbcm)|^2 +|\Scv(\bbcm)|^2,$ where
   \eqn\chargedrate{ \eqalign{ \Scu(\bbcm) &= \sum_{ij} V_{ei} V_{ej} b_{ij}
   \int {d^4\ell \over (2\pi)^4} \; { 2 \vec{\ell}^{\;2} (w_3 - 2i w_4 ) \over
   (\ell^2 + \mnu{i}^2 -i\varepsilon )(\ell^2 + \mnu{j}^2 -i\varepsilon)}, \cr
 \Scv(\bbcm) &= \sum_{ij} V_{ei} V_{ej} b_{ij}   \int {d^4\ell \over (2\pi)^4}
 \; { \ell^0 (w_2 - \ell^2  w_5 )  \over   (\ell^2 + \mnu{i}^2 -i\varepsilon
 )(\ell^2 + \mnu{j}^2 -i\varepsilon)}.\cr}} Notice that whereas $\Sca(\bbom)$
has the same combination of form factors as $\bbzn$, $\bbcm$ depends on {\it
different} nuclear matrix elements.  (For $\bbzn$, $\Sca \propto
\sum_i V_{ei}^2 \mnu{i}$ $\int [d^4\ell/(2\pi)^4] \; [(4w_1 - w_2 +\ell^2 w_5)
/ (\ell^2 + \mnu{i}^2 - i\varepsilon)]$; these are related to the Fermi-- and
Gamow-Teller--type form factors by $w_1 = \frac{1}{3} w_{\sss GT}$ and $w_2 =
w_{\sss F} + \frac{1}{3} w_{\sss GT}$.)

A striking feature of both eqs.~\ordinaryrate\ and \chargedrate\ is that they
vanish if all neutrinos share a common mass, for in this limit the sum over
states gives $\sum_{ij} V_{ei} V_{ej} a_{ij} = a_{ee}$.  The latter is zero
since the $Z^0$ width constraint implies that the Majoron cannot directly
couple
to $\nu_e$. As an important
corollary, the amplitudes must also vanish if all of the $\nu_i$ are negligibly
light compared to the Fermi momentum $\pf$, above which the integrals in
\ordinaryrate\ and \chargedrate\ are effectively cut off. Thus in {\it any}
model explaining the excess electron events by Majoron couplings to a virtual
neutrino, at least one of the $\nu_i$ must have a mass  $\mnu{i} \gsim 100$
MeV.

To make contact with the literature, we use the predictions of the GR model as
a benchmark.  In this case
      \eqn\grme{\Sca(\bbm)_{\sss GR} = {2\sqrt{2}
	\over \pi} \,\ga^2 \,\geff      \;\Scm,}
$\geff$ is the GR coupling of $\varphi$ to $\nu_e$,  $\gv$ and $\ga$ are the
axial and vector couplings of the nucleon weak currents, and $\Scm =
\VEV{[\sn\cdot\sm - (\gv/\ga)^2]h(r)}$ (see eq.~(18)).  We integrate the
phase space $d\Omega_1$ above a threshold $E_{\rm th}$ where the anomalous
events begin and the contribution from ordinary $\bb$ decay should be small.
Table 1 compares this with the excess events seen by the Irvine group for the
elements \Se, \Mo\ and \Nd\ \Moe, and in the published spectrum of \Ge\
\ref\ge{F.T.~Avignone III \etal, \pl{B256} (1991) 559.}. In all of these cases
the excess events comprise $R = 2$ to 3 \% of the total number observed.  We
also analyze the geochemically observed decays, assuming all the events are due
to $\bbm$. Remarkably, $g_{\rm eff}$ lies in the range $\SN{4}{-5}$ to
$\SN{4}{-4}$ for all seven elements.

\midinsert\noindent {\ninerm Table 1: parameters for emission
of GR Majorons in double beta decay. $T_{1/2}^{-1}$ is the inverse half-life of
the {\it anomalous} events; note that for the last three elements this is
assumed to be the entire rate; $|\M|^2$ is the $0\nu$ matrix element extracted
from Staudt \etal; and $R$ is the ratio of anomalous to all events ---
hypothesized to be 1 for the geochemically observed decays. $\Omega_1m_e^{7}$
is the total phase space for Majoron events (see eq.~\phsp), while $E_{\rm th}$
(MeV) denotes our choice for the threshold value of the sum of the electron
energies, above which essentially only excess events appear.
$\Delta\Omega_1m_e^{7}$ is the part of phase space occurring above $E_{\rm th}$
and $\geff = g_{ee}$ is the coupling needed to explain the rate using Majoron
emission in the GR model.}
$$\vbox{\tabskip=0pt \offinterlineskip
\halign to \hsize{\strut#& \vrule#\tabskip 1em plus 2em minus .5em&
\hfil#\hfil &\vrule#& \hfil#\hfil &\vrule#& \hfil#\hfil &\vrule#&
\hfil#\hfil &\vrule#& \hfil#\hfil &\vrule#& \hfil#\hfil &\vrule#&
\hfil#\hfil &\vrule#& \hfil#\hfil &\vrule#\tabskip=0pt\cr
\noalign{\hrule}
&& Element && $T^{-1}_{1/2}$(y$^{-1})$ && $|\M|^2m_e^{-2}$ && $R$ &&
$\Omega_1$ && $E_{\rm th}$ && ${\Delta\Omega_1}$ &&
$g_{\rm eff}$ &\cr
\noalign{\hrule}
&&\Ge\ && \SNt{2}{-23} && \SNt15  && 0.02 && 2.0 && 1.5 && 0.9  && \SNt{1}{-4}
&\cr
&&\Se\ && \SNt{2}{-22} && \SNt94  && 0.03 && 17  && 2.2 && 7.6  && \SNt{8}{-5}
&\cr
&&\Mo\ && \SNt{3}{-21} && \SNt24  && 0.03 && 34  && 1.9 &&  22  && \SNt{4}{-4}
&\cr
&&\Nd\ && \SNt{3}{-20} && \SNt15  && 0.02 && 260 && 2.2 && 155  && \SNt{2}{-4}
&\cr
\noalign{\hrule}
&&\Te{128}\ &&\SNt{5}{-25}&&\SNt74 && 1.00 && 0.23&& 0.0 && 0.23 &&
\SNt{4}{-5}&\cr
&&\Te{130}\ &&\SNt{1}{-21}&&\SNt54 && 1.00 && 30  && 0.0 &&  30  &&
\SNt{1}{-4}&\cr
&&\U\   && \SNt{5}{-22} && \SNt34  && 1.00 && 33  && 0.0 &&  33  &&
\SNt{2}{-4}&\cr
\noalign{\hrule}
}}$$
\endinsert

The predictions for \Te{}\ are of particular interest because of a recent
measurement of the ratio of decay rates $\zeta \equiv
\G$(\Te{130})$/\G$(\Te{128})$= (2.41\pm 0.06)\times 10^3$ \ref\Haxton{This
result of Bernatowicz and Holenberg, together with its application to a $\bbom$
description of the excess events may be found in the talk by W.~Haxton, at
Neutrino 92, Granada, Spain}. If the same coupling  needed to account for the
endpoint anomalies ($\geff\sim\SN{1}{-4}$) is used for these decays then the GR
and $\bbom$ models predict too small a ratio: $\zeta(\bbom) = (30.4/0.23)(5/7)
= 93$. The same is not true for $\bbcm$ models because the rates scale with an
extra factor of $Q^2$ relative to those for $\bbom$, and \Te{128}\ has a small
endpoint energy, $Q\approx 0.9$ MeV. Thus we find $\zeta(\bbcm) = 770$, in much
better agreement with the experimental value.

To see whether the ordinary or charged Majoron models can predict a large
enough rate, we display the combinations of couplings, defined in \yukawa, that
are equivalent to the GR model with coupling $\geff$, so far as the total rate
is concerned.  For ordinary Majorons
  \eqn\geffinbbom{ \geff = \sum_{ij} V_{ei} V_{ej}\; \left\langle{1\over\ell^2}
     \right\rangle^{-1}\left\langle{
    m_{\nu_i} a_{ij} m_{\nu_j} - \ell^2 b_{ij} \over (\ell^2 + m^2_{\nu_i})
     (\ell^2 + m^2_{\nu_j}) } \right\rangle.}
Here $\ell$ is the four-momentum of the virtual neutrino, which as indicated
by $\VEV{\;}$, is integrated weighted by the nuclear form factors. An estimate
for the size of $\geff$ is therefore obtained by replacing $\ell^2$ by the
characteristic momentum scale, $\pf^2 \sim (100$ MeV$)^2$.  For charged
Majorons, in order of magnitude
  \eqn\geffinbbcm{ \geff = -\sum_{ij} V_{ei} V_{ej}\;\left\langle{1\over\ell^2}
     \right\rangle^{-1} \left\langle { Q \ell
     b_{ij} \over (\ell^2 + m^2_{\nu_i})(\ell^2 + m^2_{\nu_j}) }
    \right\rangle \; X.}
The two qualitatively new features here are: (1) the additional suppression by
the endpoint energy, $Q/p$, due to the softer electron spectrum of these
models, and (2) a ratio, $X$, of the $\bbcm$ to $\bbom$ nuclear matrix
elements. Later we give a more quantitative comparison of these matrix
elements, in a specific model.

Consider now a representative model for each class. For $\bbom$, augment the
Standard Model with two Majorana-Dirac sterile neutrinos, $s_+$ and $s_-$, and
a complex singlet scalar, $\varphi$.   The model consists of interactions that
preserve electron-type lepton number, assuming $L_e(\Pl s_\pm) = \pm 1$, and
$L_e(\varphi) = -2$.  To satisfy the $\bbzn$ constraint we take
$\VEV{\varphi}=0$ and fine-tune  the $\varphi$ mass to be $\ltwid 1$ MeV as
required by the $\bbm$ kinematics. The spectrum contains one Dirac neutrino
$(\psir, \psil)$ with mass $M$, $\psir = s_-^c$ and  $\psil = \nu_e \sin\theta
+
s_+ \cos\theta$.   The massless orthogonal combination $\nu_e'$ is
predominantly $\nu_e$. The remaining two neutrinos $\nu_\mu$ and $\nu_\tau$
acquire Majorana masses at dimension five.

In terms of the relevant Yukawa couplings to $\varphi$,
	\eqn\yukawamodel{ \Scl_y = -\hf g_+ \, (\bar s_+ \Pl s_+) \; \varphi
       -\hf g_- \, (\bar s_- \Pl s_-) \; \varphi^* + \hbox{c.c.} }
the effective $\bbm$ coupling in this model becomes
	\eqn\geffi{\geff = \left\langle{1\over\ell^2}
     \right\rangle^{-1} \; \left\langle {(g_- \ell^2 - g_+ M^2
 	  \cos^2\!\theta) M^2\sin^2\!\theta \over
	    \ell^2 (\ell^2 + M^2)^2} \right\rangle. }
Since the light scalars couple to the massless neutrino $\nu_e'$ through
mixing, $\bbom$ may proceed via the exchange of $\nu_e'$, with the result
$\geff$ is nonvanishing as $M\to \infty$ with $\theta$ held fixed. (Decoupling
is not violated since $\theta \sim 1/M$ if dimensionless couplings in the
lagrangian are held fixed for $M$ large.)  Using $\VEV{\ell^2} \sim (100$
MeV$)^2$ and $\theta \sim 0.1$ we find $\geff\sim 10^{-4}$, as suggested by the
$\bb$ anomalies. We choose $M$ above the $K^\pm$ meson mass; otherwise a heavy
neutrino with such large mixing would have been seen as a peak in the $K\to
e\nu$ \ref\Kconstraints{R.E.~Shrock, \pr{D24} (1981) 1232; T.~Yamazaki \etal,
{\it Proceedings of the XIth International Conference on Neutrino Physics and
Astrophysics,} eds. K.~Kleinknecht and E.A.~Paschos (World Scientific,
Singapore, 1984), p.~183.} or $\pi\to e\nu$ \ref\piconstraints{D.I.~Britton
\etal, \prl{68} (1992) 3000.} decay spectra.

For charged Majorons we suppose that the lepton symmetry group surviving at
electroweak energies is $G_L = SU(2)_s \times U(1)_{L'}$, and introduce the
sterile Majorana neutrinos $N = (N_+, N_-)$, $s_+$, and $s_-$, whose
left-handed components transform respectively as $(2,0)$, $(1,1)$, and $(1,-1)$
under $G_L$. A sterile scalar doublet $\phi$ transforming as $(2,1)$ is also
added, whose VEV breaks $G_L \to U(1)_{e}$ where $L_e = 2T_3 + L'$.  The
renormalizable interactions predict three massless neutrinos $\nu_e'$,
$\nu_\mu$
and $\nu_\tau$, and two heavy Dirac neutrinos of mass $M$ and $\Lambda$, say.
$\nu_e$ is mostly $\nu_e'$, but mixes with $N_+$ and $s_+$ similarly to the
previous model. There are three massless Goldstone boson states: a complex
scalar $\varphi$ having charge $L_e(\varphi) = -2$ and a real scalar with $L_e
=
0$.

For $\bbcm$ the unbroken lepton number not only forbids contributions from
graphs involving purely massless neutrinos, but it also causes the contribution
from each heavy-neutrino line to go like $1/M^2$ rather than $1/M$ when $M >
100$ MeV. The equivalent GR coupling of the Majoron in $\bbcm$ is roughly
     \eqn\ourgeff{\geff \sim \sin^2\theta {M\over 2v} {Q \pf M^2\over
       (\pf^2+M^2)^2}X,}
if $M\sim\Lambda$ is the heavy neutrino mass and $v$ is the
$G_L$-breaking scale.  (cf.~eq.~\geffinbbcm, which due to the relation between
$b_{ij}$ and the mass matrix does reproduce \ourgeff, despite appearances.)
For $M \sim v \sim \pf$, $\theta\sim 0.1$ and $X\sim 1$, $\geff$ is of order
$\theta^2 Q/\pf\sim 10^{-4}$ as desired.

To be more quantitative, we must evaluate the new nuclear matrix elements for
$\bbcm$, which differ from the usual ones because the leptonic part of the
amplitude is proportional to the Majoron and virtual neutrino momenta through
the factor $[\ellsl,\qsl]$. The $\ell_0$ piece ($\Scv$ of eq.~\chargedrate)
vanishes in the integral $\int d\ell_0$, up to small $O(Q/p)$ corrections.  To
have a $0^+\to 0^+$ nuclear transition, the  $\ell_i$ piece ($\Scu$ in
eq.~\chargedrate) must combine with odd-parity nuclear operators, which come
from the recoil corrections to the nucleon currents and p-wave Coulomb
corrections to the electron wave functions. Although they are formally of
higher order in $\alpha$ or $v/c$ of the nucleon than the usual Gamow-Teller
and Fermi matrix elements, explicit calculations by nuclear theorists show that
they need not all be small, as we will show \ref\notei{The differences between
charged and ordinary Majorons can be seen in ref.~\Doi, in what they call the
``correction term;'' in contrast to their general analysis,  our model has part
of this correction term being the leading contribution.  Our results cannot
simply be extracted from ref.~\Doi\ because we differ on the form of the
neutrino potential, and in having the matrix element $\M_2$.}.

For small mixing angles the matrix element becomes
     \eqn\bbcmme{\left|\Sca(\bbcm)\right|^2 = {2\over\pi^2} \,(\theta\ga)^4
\left({M\over v}\right)^2\sum_{i=1,2}\left|\M_i\right|^2,}
in which the nuclear matrix elements are
     \eqn\nme{ |\M_1|^2 = {7\over9}\VEVs{(A_1+iA_2)\hti};\qquad
     |\M_2|^2= {2\over9}\VEVs{A_3 \hti},}
and $A_i$ are operators
     \eqn\ops{\eqalign{A_1 &= \hbr\cdot\bigl[(C_m\sn-C_n\sm)
     +(\gv/\ga)^2(\Dn-\Dm)\cr
     &+ i(\gv/\ga)(\Dn\times\sm+\sn\times
     \Dm)\bigr];\cr
     A_2 &= {\alpha Z r\over 2 R}
     \left[(\gv/\ga)^2 + \sm\cdot\sn - 2\sm\cdot\hbr\;\sn\cdot\hbr
     \right];\cr
     A_3 &= \hbr\cdot\left[(C_m\sn+C_n\sm)
     +(\gv/\ga)^2(\Dn+\Dm)\right].\cr}}
We use the notation $\VEV{O_nO_m} = \langle 0^+_f |\sum_{n,m} O_nO_m \tau^+_n
\tau^+_m |0^+_i\rangle$, where $\tau^+_n$ is the isospin raising operator for
the $n$th nucleon.  Definitions of $C$ and ${\bf D}$ can be found in the
reviews
by Doi \etal\ \ref\reviewD{M.~Doi, T.~Kotani and E.~Takasugi,
Prog.~Theo~Phys.~Suppl.~{\bf 83} (1985) 1.}\ or Tomoda \ref\reviewT{T.~Tomoda,
Rept.~Prog.~Phys.~{\bf 54} (1991) 53.}. The neutrino potential $\hti$ depend on
the internucleon separation vector ${\bf r}={\bf r}_n -{\bf r}_m$ and the heavy
neutrino mass $M$:
     \eqn\potential{ \hti = {\partial\over\partial M^{2}}{\partial\over\partial
  r}\left(h(0)-h(M)\right);\qquad
      h(M) =  \int{d^3p\over 2\pi^2}e^{i{\bf p}\cdot {\bf r}}
     {1\over\omega(\omega+\mu)},}
where $\omega = (p^2 + M^2)^{1/2}$ and $\mu \sim 10$ MeV is the average
nuclear excitation energy.

In Table 2 we show the value of the nuclear matrix elements,
$(\sum_i|\M_i|^2)^{1/2}$, needed for charged Majoron emission to account for
the anomalous events, assuming that $\theta^2(M/v) = 3\times 10^{-2}$
(as would be the case for a neutrino with 17\% mixing, the limit coming
from flavor universality \ref\univerality{M.~Gronau, C.N.~Leung and
J.L.~Rosner, Phys.~Rev.~{\bf D 29,} (1984) 2539.}).
For comparison, we also show a similar matrix element that has been calculated
by Muto \etal\ \ref\Muto{K.~Muto, E.~Bender and H.V.~Klapdor, Z.~Phys.~{\bf A
334} (1989) 187, as tabulated in \reviewT.}, namely
     \eqn\muto{\M_{\sss R}= (\gv/\ga)\VEV{\pf^2}^{-1} \VEV{(\Dn\times
     \sm+\sn\times \Dm)\cdot\hbr\;\; \partial h/\partial r},}
Except for a minor difference in the neutrino potential, the same matrix
element is contained in the $A_1$ term of eq.~\nme.  This is only meant to be
indicative because there are further contributions to our $\sum_i |\M_i|^2$
which do not appear in the literature, and which are conceivably bigger than
$\M_{\sss R}$.

\midinsert
{\ninerm\noindent Table 2: phase space and matrix elements for charged Majoron
emission in double beta decay.  The phase space factors are defined analogously
to those in table 1.  We give the values of the matrix elements which are
needed to explain the data (assuming all events were $\bbcm$ for \U\ and \Te{})
for $\theta^2 M/v =3\times 10^{-2},$ and values of some representative matrix
elements that have been calculated; see eqs.~\nme\ and \muto, respectively.}
$$\vbox{\tabskip=0pt \offinterlineskip
\halign to \hsize{\strut#& \vrule#\tabskip 1em plus 2em minus .5em&
\hfil#\hfil &\vrule#& \hfil#\hfil &\vrule#& \hfil#\hfil &\vrule#&
\hfil#\hfil &\vrule#& \hfil#\hfil &\vrule#\tabskip=0pt\cr
\noalign{\hrule}
&& Element && ${\Omega_3/m_e^9}$ && ${\Delta\Omega_3/m_e^9}$ &&
$\left(\sum_i |\M_i|^2\right)^{1/2}$ needed && $|\M_{\sss R}|$ (Muto \etal)&\cr
\noalign{\hrule}
&&\Ge\      && 4.7  && 0.43 && 1.1   && 1.1 &\cr
&&\Se\      && 78   && 7.9  && 0.99  && 0.95 &\cr
&&\Mo\      && 160  && 42   && 1.4   && 1.1 &\cr
&&\Nd\      && 1570 && 320  && 1.6   && 1.3 &\cr
\noalign{\hrule}
&&\Te{128}&& 0.13 && 0.13  && 0.32    && 0.92 &\cr
&&\Te{130}&& 110 && 110    && 0.48  && 0.78 &\cr
&&\U\       && 35   && 35   && 0.61  &&   ? &\cr
\noalign{\hrule}}}$$
\endinsert

 From table 2 we see that charged Majoron emission explains the data remarkably
well, if the as-yet uncomputed matrix element for \U\ is similar to those of
the other nuclei. In particular the $\bbcm$ rates for the geochemically
observed decays of \Te{128}\ and \Te{130}\ dominate over the contribution of
$\bbtn$. As was described earlier, this is in better agreement with the data
than is the result for $\bbom$ models.

Because of the extra suppression by $(Q/\pf)^2$ in $\bbcm$, we needed a heavy
neutrino $\Psi$ mixing with $\nu_e$ at the $\theta=10\%$ level, with a mass
near
100 MeV. But there are strong limits on $\theta$ for such a heavy
neutrino. Of these, beam-dump experiments do not apply to our model because
$\Psi$ decays invisibly into a light neutrino plus a Majoron.
However, searches for peaks in the spectra of
$\pi,K\to\nu e$ \ref\constraints{R.E.~Shrock, \pr{D24} (1981) 1232; T.~Yamazaki
\etal, {\it Proceedings of the XIth International Conference on Neutrino
Physics and Astrophysics,} eds. K.~Kleinknecht and E.A.~Paschos (World
Scientific, Singapore, 1984), p.~183.} rule out a long-lived heavy neutrino
with
such large mixing, unless $M>M_K \approx 500$ MeV. For such large $M$ the
$\bbcm$ rate becomes suppressed, unless the unknown nuclear matrix elements
turn out to be considerably larger than those of ref.~\Muto.

But there is a loophole in the peak search bounds: they do not apply if the
$100$ MeV neutrino $\Psi$ is a broad resonance. A width of 10 MeV is sufficient
to hide the $\pi\to\nu\Psi$ peak beneath backgounds \ref\brit{ We thank D.
Britton for helpful conversations on this point.}, and the $K$ decay searches
do not extend below 140 MeV.   Such large widths occur in our model if the
neutrino-Majoron coupling is strong, $g \sim 4\pi$, in which case we must
repeat the $\bbm$ analysis with new form factors parameterizing the
strongly-coupled neutrino sector.  Then dimensional analysis again implies that
the rate is maximal if the scale for new neutrino physics is of order 100 MeV.
It would be fascinating if double beta decay experiments gave the first hint of
a strongly-coupled neutrino sector!

We note that $v\sim M \sim$ (few hundred MeV) is also the scale at which the
global symmetry $G_L$ must break. While this is still annoyingly small compared
to the weak scale, it is a significant improvement on the much smaller 10 keV
scales needed in ordinary Majoron models.

A final challenge confronting any model is the strong limit, $N_\nu < 3.3$,
\ref\nucleo{T.P.~Walker, G.~Steigman, D.N.~Schramm, K.A.~Olive and H.S.~Kang,
Ap.~J.~{\bf 376} (1991) 51 and references therein.} on the number of neutrino
species at nucleosynthesis.  The large coupling implied by the $\bb$ anomalies
means that Majorons had an undiluted thermal energy density at MeV
temperatures, a potential disaster, since each Majoron counts as 4/7 of a
neutrino. The three Majoron states of the $\bbcm$ models therefore preclude
more than {\it two} light neutrinos at nucleosynthesis, which could happen if
$\nu_\tau$ (or $\nu_\mu$) decays before 0.8 MeV.  One possibility is that
$m_{\nu_\tau} \sim 25$ MeV (large enough to allow it to decouple before
nucleosynthesis but still below the laboratory mass bound), with a lifetime
$\tau< 10^3$ sec \ref\fermilab{E.W.~Kolb, M.S.~Turner, A.~Chakravorty and
D.N.~Schramm, \prl{67} (1991) 533.}.  But even if $\nu_\tau$ is light, it can
effectively contribute as {\it negative} number of neutrino species if it
decays into $\nu_e \varphi$ with lifetime $6\times 10^{-4}$ sec $<\tau< 2\times
10^{-2}$ sec \ref\kimmo{K.~Enqvist, K.~Kainulainen and M.~Thomson, \prl{68}
(1992) 744.}, an even better prospect for our models.  The first scenario, for
example, occurs in our ordinary Majoron model via $\nu_\tau \to \nu_\mu +
\varphi$, mediated by a dimension-seven operator $(L_\mu H) \Pl (L_\tau H)
\varphi^* \varphi$. Similarly, we can use dimension-five and dimension-seven
interactions to give majorana masses and lifetimes to $\nu_\mu$ and $\nu_\tau$
in the charged Majoron model.

In summary, we have presented two classes of models for Majoron emission in
double beta decay which may be able to explain excess events near the endpoints
of several elements without seriously spoiling the agreement of geochemical
$\bb$ observations with theory.  We propose a new class of models involving a
`charged' Majoron that carries an unbroken lepton number, which predict a
softer sum energy spectrum than that of ordinary Majorons, and thus make them
experimentally distinguishable from the latter. Charged-Majoron models are
highly constrained, and require a larger, 100 MeV scale of symmetry breaking
than the unnaturally small 10 keV scale of ordinary Majorons---a $10^4$-fold
improvement.  The detailed predictions for these models depend in part on the
size of certain nuclear matrix elements (eq.~\nme) which differ from those
appearing in the usual amplitudes for $\bbtn$, $\bbzn$, or $\bbom$. All our
models point toward a sterile neutrino with a mass of a few hundred MeV, a
large ($\sim 0.1$) angle for mixing with $\nu_e$, and possibly a large ($\sim
10$ MeV) width. Nucleosynthesis requirements suggest that the neutrino which is
predominantly $\nu_\tau$ should either have a mass that is not much below the
laboratory lower limit, or else a lifetime $\ltwid 10^{-3}$ sec.

We warmly thank F.~Avignone, M.~Moe and A.~Turkevich for information about
their experiments, T.~Kotani, R.~Shrock and E.~Takasugi for helpful
correspondence, W.~Haxton, M.~Luty, N. de Takacsy and P.~Vogel for helpful
discussions and R.~Fernholz for his contributions,including the figure. This
work was supported in part by the Natural Sciences and Engineering Research
Council of Canada and les Fonds F.C.A.R.~du Qu\'ebec.

%\vskip 0.5in
\vfill\eject
\listrefs
\end